\documentclass[twocolumn,showpacs,preprintnumbers,amsmath,amssymb]{revtex4}
 \usepackage{graphicx}% Include figure files
\usepackage{dcolumn}% Align table columns on decimal point
\usepackage{bm}% bold math

\newcommand{\newc}{\newcommand}
\newc{\simgt}{\lower.7ex\hbox{$\;\stackrel{\textstyle>}{\sim}\;$}}
\newc{\simlt}{\lower.7ex\hbox{$\;\stackrel{\textstyle<}{\sim}\;$}}

\begin{document}
 %\DeclareGraphicsExtensions{.pdf,.gif,.jpg}

\preprint{IZTECH-P2009/05, CUMQ-HEP/155}

\title{Sneutrino Dark Matter: Symmetry Protection and Cosmic Ray 
Anomalies  }

\author{Durmu\c{s} A. Demir$^{a}$}
\author{Lisa L. Everett$^{b}$}
\author{Mariana Frank$^c$}
\author{Levent Selbuz$^{a,d}$}
\author{Ismail Turan$^{c,e}$}
\affiliation{$^a$Department of Physics, Izmir Institute of
Technology, IZTECH, TR35430 Izmir, Turkey,}
\affiliation{$^b$Department of Physics, University of
Wisconsin, Madison, WI53706, USA,} \affiliation{$^c$Department
of Physics, Concordia University, 7141 Sherbrooke St. West,
Montreal, Quebec, Canada H4B 1R6,} \affiliation{$^d$Department
of Engineering Physics, Ankara University, TR06100 Ankara,
Turkey,}\affiliation{$^e$Ottawa-Carleton Institute of Physics, 
Carleton University,
1125 Colonel By Drive
Ottawa, Ontario, Canada, K1S 5B6.}

\date{\today}% It is always \today, today,

\begin{abstract}
We present an $R$-parity conserving model of sneutrino dark
matter within a Higgs-philic $U(1)^\prime$ extension of the
minimal supersymmetric standard model.  In this theory, the
$\mu$ parameter and light Dirac neutrino masses are generated
naturally upon the breaking of the $U(1)^\prime$ gauge
symmetry.  One of the right-handed senutrinos is the LSP.
The leptonic and hadronic decays of another sneutrino,
taken to be the  next-to-lightest superpartner,
allow for a natural fit to the recent results
reported by the PAMELA experiment. We perform a detailed calculation of the dark matter relic density  in this scenario, and show that the model is consistent with the ATIC and FERMI-LAT experiments. 
\end{abstract}

%Use showkeys class option if keyword
\maketitle

%%%%%%%%%%%%%%%%%%%%%%%%%%%%%%%%%%%%%%%%%%%%%%
\section{Introduction and Motivation}
%%%%%%%%%%%%%%%%%%%%%%%%%%%%%%%%%%%%%%%%%%%%%%

The question of the
nature of dark matter (DM), which forms nearly one-fourth of
the total mass in the universe, is a pivotal question for
cosmology, astrophysics, and particle physics. Other than its
relic density \cite{wmap}, little is known about the structure
of DM. Recently, there has been great excitement that new
information about DM may be revealed by the results from the
PAMELA satellite \cite{pamela} claiming an increase in positron
fraction in cosmic rays with energies above $10\ {\rm GeV}$.
Others, including the PPB-BETS \cite{ppb-bets} and HESS
\cite{hess} experiments, also claim enhancements in
electron/positron flux for energies above $100\ {\rm GeV}$.
More recently, the FERMI collaboration released a measurement of the 
$e^+e^-$ flux in the $20\ {\rm GeV}$ to $1\ {\rm TeV}$ 
range \cite{Fermi}.  All of these experiments are consistent with a 
new primary source contributing to 
local electron/positron fluxes in the $10-1000\ {\rm GeV}$ range 
\cite{Chang:2009zs}.
These anomalies in cosmic ray fluxes may well originate from
astrophysical sources such as pulsars or nearby supernova
remnants \cite{astro}. However, they might also
result from the presence of DM particles, providing a
golden opportunity to learn more about the properties of DM.

This intriguing possibility has led to a number of proposals
that interpret the data as arising from dark matter annihilation 
\cite{ann}
(which necessitates an ${\cal{O}}(100)$ boost factor) or decays
\cite{decay} (which must proceed much slower than the present
value of the Hubble parameter). Moreover, PAMELA, while
reporting an excess in the electron/positron flux, claims no
excess in the proton/anti-proton flux \cite{pamela2}. Thus, one
intriguing idea is that the DM carries lepton number, {\it e.g.},
it is composed of right-handed scalar neutrinos (sneutrinos) in
minimal \cite{moroi,pilaftsis,mcdonald} and extended
\cite{matchev,munoz,cerdeno} supersymmetric models. This framework
might explain the preference for leptonic annihilation/decay
products
\cite{takahashi,santoso,ma,gogoladze,leptophilic,leptonichiggs,maxim}.

In the simplest extensions of the minimal supersymmetric model (MSSM) 
that include right-handed neutrinos, the right-handed neutrinos are gauge singlets.   
Hence the strength of their  Yukawa interactions with the standard model (SM) 
fields governs whether they thermalize together with the rest of the MSSM matter
\cite{matchev,pilaftsis,santoso}. On the other hand, if the
lightest sneutrino $\widetilde{\nu}^1_{R}$ is the lightest supersymmetric particle (LSP)
and therefore is a potentially viable DM candidate,  it can be produced 
with the
right relic density from the decays of heavier superpartners \cite{moroi}.
% {\it i.e.}, the LSP $\widetilde{\nu}^1_R$ pops up from the thermal bath of sparticles.
If the heavier right-handed scalar neutrino $\widetilde{\nu}^2_R$ is the 
next-to-lightest superpartner (NLSP), its lifetime turns out to be longer 
than the age of the universe \cite{moroi,mcdonald}.  Therefore, decays of 
$\widetilde{\nu}^2_R$ should still be active, resulting in observable effects 
like the excess energetic electrons/positrons in cosmic rays. In the MSSM with 
right-handed neutrinos, these features have been analyzed with respect to the 
observed cosmic ray anomalies \cite{maxim}.  However, such right-handed sneutrino 
dark matter scenarios generically suffer from two aesthetic problems: {\it (i)} the 
origin of the suppressed Dirac neutrino Yukawa couplings
 %of order${\cal{O}}\left(10^{-13}\right)$ of the Dirac neutrinos
(though such small couplings are allowed on the  basis of technical naturalness
\cite{moroi,mcdonald,maxim}), and {\it (ii)} the origin of the supersymmetric $\mu$ 
parameter \cite{kimnilles}.
In this paper, we present a $U(1)^\prime$-extended MSSM model in which these aesthetic 
problems are remedied, and the desirable features of sneutrino DM, including the 
ability to account for the PAMELA
positron excess, can be correctly produced.

%%%%%%%%%%%%%%%%%%%%%%%%%%%%%%%%%%%%%%%%%%%%%%%%%%%%%%%%%%%%%%%
\section{The Higgsphilic $U(1)^\prime$-extended MSSM}
%%%%%%%%%%%%%%%%%%%%%%%%%%%%%%%%%%%%%%%%%%%%%%%%%%%%%%%%%%%%%%%

It is well known that 
the MSSM suffers from a naturalness problem due to the presence of the superpotential 
bilinear operator  $\mu \widehat{H}_u\cdot \widehat{H}_d$ \cite{kimnilles}. Though the 
mass parameter $\mu$ enters from the superpotential, it must be of the order of the mass
terms associated with the supersymmetry breaking sector.  This puzzle can be remedied 
by extending the matter and gauge structure of the MSSM, {\it e.g.} within unified and/or
 string models \cite{gut-string}.  To this end, theories with an extra $U(1)^{\prime}$ broken 
at the electroweak- to- TeV scale by SM singlets are known to be able to generate an 
appropriately sized $\mu$ parameter (see {\it e.g.} \cite{biz}).

The $U(1)^{\prime}$ symmetry can also play a crucial role in generating neutrino masses.  
The right-handed neutrino sector and the $\mu$ parameter can be correlated for both Majorana 
\cite{biz2} and Dirac masses \cite{biz3}. We assume here that lepton number is an 
accidental symmetry that is conserved at the perturbative level. Hence, the neutrinos 
are Dirac fermions, requiring Yukawa couplings of ${\cal{O}}\left(10^{-13}\right)$.
These couplings are technically natural, but an explanation for such a strong suppression 
is clearly desirable.  One way this can occur is if the $U(1)^{\prime}$ invariance suppresses 
leading order contributions to Dirac neutrino masses and allows higher-dimensional
 operators \cite{biz3}.

In this work, we assume that the $U(1)^\prime$ charges satisfy 
$Q^{\prime}_{H_u} + Q^{\prime}_{H_d} \neq 0$ to forbid the bare $\mu$ term, 
and $Q^{\prime}_{H_u} + Q^{\prime}_{L} + Q^{\prime}_{N} \neq 0$ to forbid a 
bare neutrino Yukawa coupling. After including an SM-singlet chiral superfield 
$\widehat{S}$, the relevant part of the superpotential takes the form \cite{biz,biz3}
\begin{eqnarray}
\label{superpot}
\widehat{W} = h_{\mu} \widehat{S} \widehat{H}_u \cdot \widehat{H}_d
%+ \bar{h}_s \widehat{S}_1 \widehat{S}_2 \widehat{S}_3 \nonumber\\
%&+& \widehat{Q} \cdot \widehat{H}_u {\bf Y_U} \widehat{U}
%+ \widehat{Q} \cdot \widehat{H}_d {\bf Y_D} \widehat{D}\nonumber\\
%&+& \widehat{L} \cdot \widehat{H}_d {\bf Y_E} \widehat{E} +
+ \frac{1}{M_R} \widehat{S} \widehat{L} \cdot \widehat{H}_u {\bf
Y_{\nu}} \widehat{N}
\end{eqnarray}
in which the  $U(1)^{\prime}$  invariance requires
$Q^{\prime}_{H_u} + Q^{\prime}_{H_d} + Q^{\prime}_S = 0$ and
$Q^{\prime}_{H_u} + Q^{\prime}_{L} + Q^{\prime}_{N} +
Q^{\prime}_S =0$. In the above, $M_R$ is a large mass scale,
and $h_{\mu}$ and ${\bf Y_{\nu}}$ are the Yukawa couplings
responsible for generating the $\mu$ parameter and neutrino
masses.
%, and
%$\Delta \widehat{W}$ stands for the sector needed to cancel the
%gauge and gravitational anomalies.
Upon the breaking of the electroweak and $U(1)^{\prime}$ gauge
symmetries, the effective low energy parameters include the
$\mu$ parameter
\begin{eqnarray}
\mu = h_{\mu} \langle S \rangle ,
\label{mueff}
\end{eqnarray}
and Dirac neutrino masses
\begin{eqnarray}
\label{mneut}
{\bf m_{\nu}} = \frac{1}{M_R} \langle S \rangle \langle
H_u^0\rangle {\bf Y_{\nu}}\equiv \overline{{\bf Y}}_{\nu} \left(\langle
H_u^0\rangle/\sin\beta\right).
\end{eqnarray}
The effective neutrino Yukawa coupling $\overline{{\bf
Y}}_{\nu}$ leads to neutrino masses in good agreement with experiment:
\begin{eqnarray}
\label{hnu-eq}
\left|\overline{{\bf Y}}_{\nu}\right|
%\equiv \frac{\langle S \rangle}{M_R} \left|{\bf Y_{\nu}}\right| \sin \beta
\simeq 3 \times 10^{-13} \left(\frac{m_{\nu}^2}{2.8\times 10^{-3}\ {\rm eV}^2}\right)^{1/2},
\end{eqnarray}
for $\left|{\bf Y_{\nu}}\right| \sim 1$, $\tan\beta \equiv
\langle H_u^0\rangle/\langle H_d^0\rangle \sim 1$, and
\begin{eqnarray}
\label{param1}
\langle S \rangle \simeq 3\ {\rm TeV}\,,\; M_R \simeq M_{GUT}= 10^{16}\ {\rm GeV}\,.
\end{eqnarray}
Hence, in this framework the $\mu$ problem is resolved and appropriately suppressed
Dirac neutrino masses are generated upon
$U(1)^{\prime}$ breaking. Parameter values were chosen to obtain 
neutrino masses which are in the right range. In the next sections, we give more details of the spectrum   constrained by the experimental data.

%%%%%%%%%%%%%%%%%%%%%%%%%%%%%%%%%%%%%%%%%%
\section{Cosmic Ray Anomalies} 
%%%%%%%%%%%%%%%%%%%%%%%%%%%%%%%%%%%%%%%%%

To account for the
anomalous cosmic ray fluxes, there arise several additional
constraints on the model. First, we assume the right-handed neutrino
superfields are {\it total gauge singlets} to avoid unacceptably fast decays 
that would otherwise occur via gaugino mediated processes.
 %({\it e.g.} via exchanges of the  $U(1)^{\prime}$ gaugino $\widetilde{Z}^{\prime}$).
  Second, the $U(1)^{\prime}$ charges must be assigned such that the successful
  generation of the $\mu$ parameter and neutrino masses are maintained.
  The option that we pursue in this work is to have a ``Higgs-philic" $U(1)^{\prime}$, 
similar in spirit to \cite{leptonichiggs}.
%  Finally, the gauge and gravitational anomalies must be cancelled, necessitating the addition of
%  exotic matter (or family-dependent charges, which we will not study in this work).
%  %The new states must not spoil the success of the model in  both the cosmological and 
%particle physics contexts.

Enforcing these constraints leads to the charge assignments
displayed in Table \ref{charges}. Note that of the MSSM fields, only the
right-handed up quarks $\widehat{U}$ and the Higgs fields
$\widehat{H}_u$ and $\widehat{S}$ have nonvanishing
$U(1)^{\prime}$ charges. %As the bilinears $\widehat{S}
%\widehat{H}_u$ and $\widehat{S} \widehat{U}$ are
%forbidden to appear in the superpotential by SM gauge invariance, the bare $\mu$ term does 
%not get regenerated.
The charge assignment in Table \ref{charges} is anomalous as it
stands. However, all gauge and gravitational anomalies can be
cancelled either by invoking family-dependent $U(1)^{\prime}$ charges as in \cite{kane} 
(though in this case one needs to worry about constraints from flavor violation), or by 
augmenting the matter content of the theory by sets of vector-like quark and lepton fields, 
as well as additional SM singlets that are charged under the $U(1)^\prime$ gauge symmetry. 
The details will be addressed somewhere else in our future studies.

With this charge assignment, scattering processes that involve $\widehat{H}_u$, $\widehat{S}$ 
and $\widehat{U}$ are influenced by the $U(1)^{\prime}$ gauge boson ($Z^{\prime}_{\mu}$) and the
$U(1)^{\prime}$ gaugino ($\widetilde{Z}^{\prime}$) \cite{ali}. At hadron colliders, the 
right-handed up-type quarks ($u_R$ and $c_R$) can undergo Drell-Yan annihilation through 
$Z^{\prime}_{\mu}$ to produce Higgs fields. The decays of the up-type squarks also exhibit 
novel branchings due to $\widetilde{Z}^{\prime}$ exchange \cite{ali}.

\begin{table}[htb]
\small\addtolength{\tabcolsep}{9pt}
\begin{tabular}{ccccc}
\hline \hline \\[-1.4ex]
$\mbox{Field}$ & $SU(3)_C$ & $SU(2)_L$ & $U(1)_Y$ & $U(1)^{\prime}$  \\[-2.0ex]\\
 \hline \hline \\[-1.4ex]
 $\widehat{Q}$ & $3$ & $2$ & $1/6$ & $0$ \\
\\[-1.4ex]
 $\widehat{U}$ & $\overline{3}$ & $1$ & $-2/3$ & $-Q_{H_u}^{\prime}$ \\
\\[-1.4ex]
$\widehat{D}$ & $\overline{3}$ & $1$ & $1/3$ & $0$ \\
\hline\\[-1.4ex]
$\widehat{L}$ & $1$ & $2$ & $-1/2$ & $0$\\
\\[-1.4ex]
$\widehat{N}$ & $1$ & $1$ & $0$ & $0$ \\
\\[-1.4ex]
$\widehat{E}$ & $1$ & $1$ & $1$ & $0$\\
 \hline\\[-1.4ex]
$\widehat{H}_u$ & $1$ & $2$ & $1/2$ & $Q_{H_u}^{\prime}$ \\
\\[-1.4ex]
$\widehat{H}_d$ & $1$ & $2$ & $-1/2$ & $0$ \\
\\[-1.4ex]
$\widehat{S}$ & $1$ & $1$ & $0$ & $-Q_{H_u}^{\prime}$ \\[-2.0ex]\\
\hline\hline
\end{tabular}
\caption{\sl\small The quantum numbers of quark ($\widehat{Q},
\widehat{U}, \widehat{D}$), lepton ($\widehat{L}, \widehat{N},
\widehat{E}$), and Higgs ($\widehat{H}_u, \widehat{H}_d$,
$\widehat{S}$) superfields. The superpotential couplings of
quarks and charged leptons are kept as in the MSSM.}
\label{charges}
\end{table}

We now explore the possibility that the excess positron flux observed in the cosmic ray data is due to
the presence of sneutrino DM. To do so, we will assume that the soft supersymmetry breaking
sector of the theory is such that the lightest right-handed sneutrino $\widetilde{\nu}^1_{R}$ 
is the LSP and the next-to-lightest right-handed  sneutrino $\widetilde{\nu}^2_R$
is the NLSP. More explicitly,
\begin{eqnarray}
m_{\widetilde{\nu}^1_R} < m_{\widetilde{\nu}^2_R} < m_{\widetilde{rest}}
\end{eqnarray}
where $m_{\widetilde{rest}}$ denotes the remaining superpartner masses, including the heaviest right-handed
sneutrino, $\widetilde{\nu}^{3}_{R}$. The third heavy particle,  the NNLSP, is  chosen to be 
the lightest neutralino, $\widetilde{\chi}_1^0$ and it requires special attention since it decays late enough to have some non-thermal contribution to the DM relic density. The hadronic Big Bang Nucleosynthesis (BBN) constraints will also be considered for the late decaying 
particles in the model. We discuss these issues further in the next section.

%%%%%%%%%%%%%%%%%%%%%%%%%%%%%%%%%%%%%%%%
\subsection{The DM relic density}
%%%%%%%%%%%%%%%%%%%%%%%%%%%%%%%%%%%%%%%%

%The LSP ($\widetilde{\nu}^1_{R}$) is the DM candidate, 
%and thus, 
For the  spectrum considered above, the DM has two components, 
the stable LSP and the practically stable NLSP which lives longer than the age of the 
universe.  Together, their relic density should reproduce the WMAP \cite{wmap} result.  
The right-handed sneutrino population forms during the cosmic evolution from the decays of the MSSM spectrum plus exotics, which, with a few exceptions,  should already have reached thermal equilibrium. A given sparticle decays into right-handed sneutrinos with a rate $\Gamma \sim
\left|\overline{{\bf Y}}_{\nu}\right|^2 \times \left(\it{\mbox{sparticle mass}}\right)$. Therefore, the
conversion rate into right-handed sneutrinos is quadratic in the effective neutrino Yukawa coupling $\overline{{\bf Y}}_{\nu}$, and the growth of the LSP number density is fast enough to produce the observed DM relic density \cite{moroi}.

We expect the third sneutrino to also decay to the LSP at freeze-out and thus contribute to 
the relic density. However, in our chosen parameter space, this sneutrino is taken very heavy (quantified below),  much heavier than other sparticles. It would likely mix more with the left-handed sneutrino and 
decay very quickly compared to the NLSP. Our detailed analysis shows that its life-time 
($\sim 10^{-5}$ sec) is larger than its freeze-out time ($\sim 10^{-10}$ sec) but its non-thermal contribution to the relic density is negligible 
since the branching ratio for decaying into LSP or NLSP is suppressed (less than $10^{-26}$). 

The right-handed sneutrinos exhibit a non-thermal distribution.  To see this, recall that 
%that while the 
%right-handed neutrino is a gauge singlet and thus is sterile, 
the right-handed sneutrinos mix with the left-handed sneutrinos and therefore interact with the gauge sector. For right-handed sneutrino annihilation, only a four-point interaction, a left-handed sneutrino, 
or a Higgsino exchange can contribute. The annihilation is out of equilibrium as long as the interaction rate is smaller than
the expansion rate $h\sim {T^2}/{M_{Pl}}$.
%CHANGE
The rate of the four-point interaction $\Gamma_4 \sim \left|{\overline{\bf
Y}}_{\nu}\right|^4 T <h $, since  $|\overline{{\bf Y}}_{\nu}|\sim 10^{-13}$.
% is consistent with  $T>10^{-33}\ {\rm
%GeV}$ for $\overline{{\bf Y}}_{\nu}\sim 10^{-13}\ $.
The left-handed sneutrino and Higgsino
exchanges also have rates $\Gamma_{{\widetilde
\nu}_L},\,\Gamma_{{\widetilde H}_u} < h$, and hence the right-handed
sneutrinos do not thermalize before the electroweak phase
transition. This is maintained after the phase
transition if the mixing with the active (left-handed)
sneutrino is small \cite{moroi,russell}.
%The right-handed sneutrinos
 %a dark matter relic density that matches the observed value
%\cite{moroi,russell}.

For a detailed calculation of the relic density, the model must be considered 
in detail and all the mixings, masses and branching ratios must be determined so that we would  
calculate the relic density by counting all possible channels. Moreover, we need to evaluate 
the would-be relic density of the late decaying NNLSP which is computed with through conventional methods. We implement the model fully 
into the {\tt CalcHEP}  \cite{calchep} package program with the help of {\tt LanHEP} \cite{Semenov:2008jy} program. Once the {\tt CalcHEP} model files are provided, we use the  {\tt MicrOMEGAs} \cite{Belanger:2008sj} software for calculating the NNLSP relic 
density. 

In the implementation, we consider a normal hierarchy for the neutrinos and chose the mixing parameters to be $\sin^2 2\theta_{12}=0.87$, $\sin^2 2\theta_{23}=0.92$, and $\sin^2 2\theta_{13}=0.02$. 
The parametrization of the mixing matrix is identical to the one in the quark sector and the additional CP-violating phases are taken zero. As the positron excess is proportional to the neutrino masses, we keep these non-zero, consistent with the following constraints: 
$\Delta m_{21}^2=(7.59\pm0.20)\times 10^{-5} {\rm eV}^2$ and $\Delta m_{32}^2=(2.43\pm0.13)\times 10^{-3} {\rm eV}^2$ \cite{Aliu:2004sq}. 
We take $M_R$, $\mu$, $\tan\beta$, and $h_\mu$ as free parameters and express 
$\langle S \rangle$, ${\bf Y}_{\nu}$ and $\overline{{\bf Y}}_{\nu}$ in terms of them.

In this model the sneutrinos also mix, and the mixing matrix 
can in general be expressed as
\begin{eqnarray}
\label{sneutrino_mix}
 {\cal L}_m^{\widetilde \nu }= -\sum_{i,j=1}^3
({\widetilde \nu}_L^{i*} 
{\widetilde \nu}_R^{j*})\left(
 \begin{array}{cc}
 m_{\widetilde \nu_{L L}^i}^{2}& m_{\widetilde \nu_{L R}^{ij}}^{2}\\[1.ex]
 m_{\widetilde \nu_{R L}^{ij}}^{2} & m_{\widetilde \nu_{R R}^j}^{2}
 \end{array}
 \right)
 \left(
\begin{array}{c}
\widetilde \nu_L^i\\ [1.ex] \widetilde \nu_R^j
\end{array}
\right),
\end{eqnarray}
where $i,j$ are the flavor indices and the matrix elements are given by
\begin{eqnarray}\label{mix_comp}
m^2_{\widetilde \nu_{LL}^i}&=&\frac{1}{4}g^2(\langle H_d^0\rangle^2-\langle H_u^0\rangle^2)
+g_Y^2(\langle H_d^0\rangle^2Y_{H_d}Y_L\nonumber \\
&&+\langle H_u^0\rangle^2Y_{H_u}Y_L)+g_{Y'}^2(\langle H_d^0\rangle^2 Q'_{H_d}Q'_L \nonumber \\
&&+\langle H_u^0\rangle^2 Q'_{H_u}Q'_L+\langle S\rangle^2Q'_LQ'_{S})\nonumber \\
&&+({\bf m}_\nu^{ii})^2+ M^2_{L_i}\nonumber \\
m^2_{\widetilde \nu_{RR}^j}&=& g_Y^2(\langle H_d^0\rangle^2Y_{H_d}Y_N+
\langle H_u^0\rangle^2Y_{H_u}Y_N)\nonumber \\
&&+g_{Y'}^2(\langle H_d^0\rangle^2Q'_{H_d}Q'_N+
\langle H_u^0\rangle^2Q'_{H_u}Q'_N \nonumber \\
&&+\langle S\rangle^2Q'_NQ'_{S})+({\bf m}_\nu^{jj})^2+ M^2_{N_j}      \nonumber\\
m^2_{\widetilde\nu_{LR}^{ij}}&=&(m^2_{\widetilde \nu_{RL}^{ij}})^*\nonumber\\
&=&{\bf m}_\nu^{ij}\left[A^*_{\nu_i} + 
\frac{\mu}{\tan\beta} 
\left(1-\left(\frac{\langle H_u^0 \rangle}{\langle S \rangle}\right)^2\right)\right].
\end{eqnarray}
Here $ M^2_{L_i}$ and $ M^2_{N_i}$ are the soft mass terms and 
$A_{\nu_i}$ are the trilinear couplings (assumed diagonal). 
${\bf m}_\nu$ and $\mu$ are given in (\ref{mneut}) and (\ref{mueff}), 
respectively. Note that due to the specific $U(1)^\prime$ charge 
assignments in Table \ref{charges}, some of the terms in 
(\ref{mix_comp}) are zero. 
The sneutrino mass eigenstates $\widetilde \nu_{1,2}^i$ are given by
\begin{eqnarray}
\begin{pmatrix}
 \widetilde \nu_1^i\\
\widetilde \nu_2^i
 \end{pmatrix} =\left(
 \begin{array}{cc}
 \cos\Theta^i_{\widetilde\nu_L-\widetilde\nu_R}& \sin\Theta^i_{\widetilde\nu_L-\widetilde\nu_R}\\[1.ex]
 -\sin\Theta^i_{\widetilde\nu_L-\widetilde\nu_R} & \cos\Theta^i_{\widetilde\nu_L-\widetilde\nu_R}
 \end{array}
 \right) \begin{pmatrix}
 \widetilde \nu_L^i\\
 \widetilde \nu_R^i
 \end{pmatrix} \quad
\label{higgs458}
\end{eqnarray}
with the left-right sneutrino mixing angles  
$\Theta^i_{\widetilde\nu_L-\widetilde\nu_R}$:
\begin{equation}\label{sneutrino_angle}
 \Theta^i_{\widetilde\nu_L-\widetilde\nu_R}
=\frac{1}{2}\arctan{\left(\frac{m_{\widetilde \nu_{R L}^{ii}}^{2}
+m_{\widetilde \nu_{L R}^{ii}}^{2}}
 {m_{\widetilde \nu_{L L}^i}^{2}-m_{\widetilde \nu_{R R}^i}^{2}}\right)},
 \end{equation}
such that $m_{\widetilde \nu_1^i}<m_{\widetilde \nu_2^i},\,i=1,2,3$. In the rest of the paper we will  
refer to the right-handed sneutrinos as $\widetilde{\nu}^i_{R},\; i=1,2,3$ for simplicity,  
but in the numerical analysis all the mixings are implemented. Note that even though 
such a mixing is small numerically, it must be retained. The alternative is to introduce 
mass insertions to obtain most of the decay channels discussed below and {\tt CalcHEP} cannot handle such insertions.
\begin{table}[b]
\small\addtolength{\tabcolsep}{8pt}
\begin{tabular}{cccccc}
\hline \hline \\[-1.4ex]
$m_{\widetilde \nu_e^1}$ & $m_{\widetilde \nu_e^2}$ & $m_{\widetilde \nu_\mu^1}$ & $m_{\widetilde \nu_\mu^2}$ & $m_{\widetilde \nu_\tau^1}$ & $m_{\widetilde \nu_\tau^2}$  \\[-2.0ex]\\
 \\[-1.4ex]
 100 & 2000 & 1200 & 2000 & 2000 & 5000 \\
\\[-1.4ex]
\hline\\[-1.4ex]
$m_{\widetilde {\chi}^0_1}$ & $m_{\widetilde {\chi}^0_2}$ & $m_{\widetilde {\chi}^0_3}$ & $m_{\widetilde {\chi}^0_4}$ & $m_{\widetilde {\chi}^0_5}$ & $m_{\widetilde {\chi}^0_6}$  \\[-2.0ex]\\
 \\[-1.4ex]
 1392 & 1519 & 1561 & 1650 & 2537 & 2538 \\
\\[-1.4ex]
\hline\\[-1.4ex]
$m_{\widetilde {\chi}^+_1}$ & $m_{\widetilde {\chi}^+_2}$ & $m_{H_1^0}$ & $m_{H_2^0}$ & $m_{H_3^0}$ & $m_{Z^\prime}$ \\[-2.0ex]\\
 \\[-1.4ex]
 1513 & 1648 & 87 & 2556 & 5410 & 2536 \\
\\[-1.4ex]
\hline\\[-1.4ex]
$m_{\widetilde e_L}$ & $m_{\widetilde e_R}$ & $m_{\widetilde \mu_L}$ & $m_{\widetilde \mu_R}$ & $m_{\widetilde \tau_L}$ & $m_{\widetilde \tau_R}$ \\[-2.0ex]\\
 \\[-1.4ex]
 1999 & 1999 & 1999 & 1999 & 1999 & 1999 \\
\\[-1.4ex]\hline \hline
\end{tabular}
\caption{\sl\small The relevant masses  in GeV for the chosen parameter set. 
$m_{\widetilde {\chi}^0_a}$ ($m_{\widetilde {\chi}^+_b}$) denotes the mass of the $a^{th} (b^{th})$ neutralino (chargino) state. 
%$H_1^0, H_2^0,$ and $H_3^0$ 
$H_{1,2,3}^0$ are the physical CP-even neutral Higgs bosons.
$m_{\widetilde \nu_e^1}$ and $m_{\widetilde \nu_\mu^1}$ 
denote the masses of the LSP and NLSP refereed as 
$m_{\widetilde \nu_R^1}$ and $m_{\widetilde \nu_R^2}$ in the text.}
\label{spectrum}
\end{table}

\begin{table}[t]
\small\addtolength{\tabcolsep}{-3pt}
\begin{tabular}{cccccc}
\hline \hline \\[-1.4ex]
$\Gamma_{\widetilde \nu_e^1}$ & $\Gamma_{\widetilde \nu_e^2}$ & $\Gamma_{\widetilde \nu_\mu^1}$ & $\Gamma_{\widetilde \nu_\mu^2}$ & $\Gamma_{\widetilde \nu_\tau^1}$ & $\Gamma_{\widetilde \nu_\tau^2}$  \\[-2.0ex]\\
 \\[-1.4ex]
 0 & 3.95 & $3.5\times10^{-52}$ & 3.95 & 3.95 & $6.5\times10^{-20}$ \\
\\[-1.4ex]
\hline\\[-1.4ex]
$\Gamma_{\widetilde {\chi}^0_1}$ & $\Gamma_{\widetilde {\chi}^0_2}$ & $\Gamma_{\widetilde {\chi}^0_3}$ & $\Gamma_{\widetilde {\chi}^0_4}$ & $\Gamma_{\widetilde {\chi}^0_5}$ & $\Gamma_{\widetilde {\chi}^0_6}$  \\[-2.0ex]\\
 \\[-1.4ex]
 $7.7\times10^{-25}$ & 0.30 & 1.0 & 1.5 & 6.2 & 11.6 \\
\\[-1.4ex]
\hline\\[-1.4ex]
$\Gamma_{\widetilde {\chi}^+_1}$ & $\Gamma_{\widetilde {\chi}^+_2}$ & $\Gamma_{H_1^0}$ & $\Gamma_{H_2^0}$ & $\Gamma_{H_3^0}$ & $\Gamma_{Z^\prime}$ \\[-2.0ex]\\
 \\[-1.4ex]
 0.13  & 1.9 & 0.0041 & 24.6 & 27176.5 & 63.7 \\
\\[-1.4ex]
\hline\\[-1.4ex]
$\Gamma_{\widetilde e_L}$ & $\Gamma_{\widetilde e_R}$ & $\Gamma_{\widetilde \mu_L}$ & $\Gamma_{\widetilde \mu_R}$ & $\Gamma_{\widetilde \tau_1}$ & $\Gamma_{\widetilde \tau_2}$ \\[-2.0ex]\\
 \\[-1.4ex]
 3.9 & $1.1\times10^{-10}$ & 3.9 & $4.6\times10^{-6}$ & 2.2 & 1.8 \\
\\[-1.4ex]\hline \hline
\end{tabular}
\caption{\sl\small The relevant total decay widths  for the particles 
in Table \ref{spectrum} in GeV for the chosen parameter set. 
$\Gamma_{\widetilde \nu_e^1}$ and $\Gamma_{\widetilde \nu_\mu^1}$ 
denote the decay widths of the LSP and NLSP.}
\label{widths}
\end{table}

For the numerical analysis, we use the following input values:
\begin{eqnarray}
 \mu=1560\,{\rm GeV},\;\; \tan\beta=0.15,\;\; M_R=10^{16}\,{\rm GeV}\nonumber\\
h_\mu=0.4,\;\; M_{L_i}=2000\,{\rm GeV},\;\; A_{\nu_i}=200\, {\rm GeV}\nonumber\\
M_{N_1}=100\,{\rm GeV},\;\; M_{N_2}=1.2\,{\rm TeV},\;\; M_{N_3}=5\,{\rm TeV}\nonumber\\
 M_1=1400\,{\rm GeV},\;\; M_2=1600\,{\rm GeV},\;\; M_3=2\,{\rm TeV}
\label{inputs}
\end{eqnarray}
where $ M_1,  M_2$ and $ M_3$ are the soft gaugino mass terms for $U(1)$, $SU(2)_L$ and $SU(3)$, 
respectively. For the chosen point in the parameter space,  the masses entering the 
calculations are given in Table \ref{spectrum}. Clearly, the LSP is $\widetilde \nu_e^1$  
and the NLSP is $\widetilde \nu_\mu^1$,  
denoted as $\widetilde \nu_R^1$ and $\widetilde \nu_R^2$ 
in the rest of the text. The third right-handed sneutrino 
$\widetilde \nu_R^3$ becomes $\widetilde \nu_\tau^2$. The NNLSP is $\widetilde \chi^0_1$
 as previously chosen. The total widths of these particles are also given in Table \ref{widths}.
 
We first  
address the hadronic BBN constraints on several particles in the table, as
 it is known that late decaying particles may spoil the predictions of BBN and thus need special attention. 
From Table  \ref{widths}, there are only two particles, namely $\widetilde \chi_1^0$ and 
$\widetilde \nu_\tau^2$, with lifetimes $\sim 1.2$ sec and $\sim 10^{-5}$ sec, respectively, which 
survive and decay after the freeze-out of the NNLSP. In our scenario, the relevant quantity \cite{Feng:2004mt} that needs to be evaluated and compared with 
the observational constraints on light elements like $\rm ^4He$, $\rm D$, $\rm ^6Li$ etc is
\begin{eqnarray}
\eta = \left(\epsilon_{\widetilde \nu_R^1} {\cal B}(\widetilde \chi_1^0\to \widetilde \nu_R^1 q\bar q X_1) + 
\epsilon_{\widetilde \nu_R^2} {\cal B}(\widetilde \chi_1^0\to \widetilde \nu_R^2 q\bar q X_2)\right)
Y_{\widetilde \chi_1^0}
\nonumber
\end{eqnarray}
where $\epsilon_{\widetilde \nu_R^{1,2}}$ are the energies carried by the hadrons and 
$Y_{\widetilde \chi_1^0}$ is the yield variable, defined as 
$Y_{\widetilde \chi_1^0}=n_{\widetilde \chi_1^0}/s$  with $n_{\widetilde \chi_1^0}$ the number 
density of $\widetilde \chi_1^0$ and $s$ the total entropy density. This quantity is calculated 
with the help of {\tt MicrOMEGAs}. We also generated all the relevant three-body decays  
eventually leading to 
hadronic final states. These are $\widetilde \chi_1^0\to \widetilde \nu_R^{1,2} \bar \nu_i H_1^0(Z) $ 
and $\widetilde \chi_1^0\to \widetilde \nu_R^{1,2} \ell^+_i W$. Then $H_1^0, Z$, and $W$ decay 
hadronically. We found that ${\cal B}(H_1^0\to q\bar q)=0.95$, ${\cal B}(Z\to q\bar q)=0.72$, and 
${\cal B}(W\to q\bar q^\prime)=0.67$. When all of the quantities are entered numerically, the variable 
$\eta$ is around $10^{-12}$ GeV.  In the range of the lifetimes of particles, 
the constraint coming from the overproduction of $^4$He is relevant. The observational 
constraint requires  $\eta({}^4\rm{He})<10^{-9.47}$ \cite{moroi}, which is satisfied for the $\widetilde \chi_1^0$ 
case. We repeated the analysis for the heaviest right-handed sneutrino 
($\widetilde \nu_\tau^2$) and  found $\eta$ even smaller than $10^{-12}$ GeV.

After insuring that the hadronic BBN constraints are satisfied, we proceed with the relic density calculation. 
As mentioned earlier, there are two particles, $\widetilde \nu_R^1$ and $\widetilde \nu_R^2$
 as the components of the DM. Including the contributions after freeze-out, the total relic 
density of the right-handed sneutrinos is given
\begin{eqnarray}
 \Omega_{\widetilde \nu_R} = \Omega^{\tt ce}_{\widetilde \nu_R^1} + \Omega^{\tt ce}_{\widetilde \nu_R^2} 
+  \Omega^{\tt fo}_{\widetilde \nu_R}
\end{eqnarray}
where ``{\tt ce}'' (``{\tt fo}'') refers the contributions from decays at chemical 
equilibrium (after freeze-out). 
The contribution to the relic density from the heaviest right-handed sneutrino is small 
and neglected here. Thus, $\Omega^{\tt fo}_{\widetilde \nu_R}$ has contributions only from the NNLSP 
  $\widetilde \chi_1^0$. It is defined as
\begin{eqnarray}
 \frac{\Omega^{\tt fo}_{\widetilde \nu_R}}{\Omega_{\widetilde \chi_1^0}} = 
 \frac{m_{\widetilde \nu_R^1}}{m_{\widetilde \chi_1^0}} 
{\cal B}(\widetilde \chi_1^0\to \widetilde \nu_R^1 \, \bar \nu_i) 
+ \frac{m_{\widetilde \nu_R^2}}{m_{\widetilde \chi_1^0}} 
{\cal B}(\widetilde \chi_1^0 \to \widetilde \nu_R^2 
 \,\bar \nu_j) \nonumber 
\end{eqnarray}
where $i,j$=1,2,3 (we also include the conjugated states). Of course,  the 
sum of the branching ratios has to be unity, {\it i.e.}, 
$\sum_{i,j=1}^3\left( {\cal B}(\widetilde \chi_1^0\to \widetilde \nu_R^1 \, \bar \nu_i) + 
{\cal B}(\widetilde \chi_1^0\to \widetilde \nu_R^2 \, \bar \nu_j)\right)=1$. Numerically, we found
$\sum_{i,j=1}^3 {\cal B}(\widetilde \chi_1^0\to \widetilde \nu_R^1 \, \bar \nu_i)=0.86$ and
 $\sum_{i,j=1}^3 {\cal B}(\widetilde \chi_1^0\to \widetilde \nu_R^2 \, \bar \nu_i)=0.14$ in the 
parameter set (\ref{inputs}). 
$\Omega_{\widetilde \chi_1^0}$ is the would-be relic density of NNLSP for the case where it is stable. 
We have calculated $\Omega_{\widetilde \chi_1^0}$ by using the {\tt MicrOMEGAs} \cite{Belanger:2008sj} 
through its {\tt CalcHEP} interface. Numerically,  we found $\Omega_{\widetilde \chi_1^0} h^2=0.09$  
with the usual dimensionless parameter $x_F\equiv T_F/m_{\widetilde \chi_1^0}=1/29.8$. Here $h$ and $T_F$ are the Hubble 
parameter and the freeze-out temperature. For example, the two most significant contributions to 
$1/\Omega_{\widetilde \chi_1^0}$ from the (co-)annihilation channels are 
$\widetilde \chi_1^0 \widetilde \chi_1^0\to t\,\bar t$ $(42\%)$ and 
$\widetilde \chi_1^+ \widetilde \chi_1^0\to t\,\bar b$ $(23\%)$. 
\begin{table}[t]
\small\addtolength{\tabcolsep}{6.5pt}
\begin{tabular}{ccccc}
\hline \hline \\[-1.4ex]
$\Omega_{\widetilde \nu_R^1}^{\tt ce} h^2$ & $\Omega_{\widetilde \nu_R^2}^{\tt ce} h^2$ & 
 $\Omega_{\widetilde \nu_R^1}^{\tt fo} h^2$ & $\Omega_{\widetilde \nu_R^2}^{\tt fo} h^2$ & 
$=\,\Omega_{\widetilde \nu_R} h^2$ \\[-2.0ex]\\
 \hline\\[-1.4ex]
 0.000112 & 0.0444 & 0.000929 & 0.0665 & $=\;0.1119$ \\ 
\\[-1.4ex]\hline \hline
\end{tabular}
\caption{\sl\small The individual contributions to the relic density for the parameter set 
 (\ref{inputs}).}
\label{relics}
\end{table}

For the calculation of $\Omega^{\tt ce}_{\widetilde \nu_R^1}$ and  $\Omega^{\tt ce}_{\widetilde \nu_R^2}$, 
the following decay channels are included (conjugated states are not listed):
\begin{eqnarray}
&&\widetilde \chi_a^0 \to \bar\nu_i \widetilde \nu_R^{1,2}\,,\;\; a=1,...6,\;\; i=1,2,3\nonumber\\
&&\widetilde \chi_b^+ \to \ell \widetilde \nu_R^{1,2}\,,\;\; b=1,2,\;\; \ell=e,\mu,\tau\nonumber\\
&&\widetilde \ell_L^+ \to W^+ \widetilde \nu_R^{1,2}\,,\;\; \widetilde \ell_L=\widetilde e_L,\widetilde \mu_L,
\widetilde \tau_1, \widetilde \tau_2\nonumber\\
&&\widetilde \nu_\mu^2 \to H_1^0 \widetilde \nu_R^{1,2}\,,\; 
\widetilde \nu_\tau^1 \to H_1^0 \widetilde \nu_R^{1,2}\,,\nonumber\\
&&\widetilde \nu_e^2 \to H_1^0 \widetilde \nu_R^2\,,\;\; \widetilde \nu_\mu^2 \to Z \widetilde \nu_R^2.
\label{channels}
\end{eqnarray}

We performed a complete parameter scan, and found these are the only decay modes which are numerically significant. The individual 
contributions to the relic density are summarized in Table \ref{relics}. In the table, we split the 
freeze-out contribution $\Omega_{\widetilde \nu_R}^{\tt fo}$ into two parts, 
$\Omega_{\widetilde \nu_R^1}^{\tt fo}$  and $\Omega_{\widetilde \nu_R^2}^{\tt fo}$, 
which show the contributions 
of the NNLSP decaying to both the LSP and the NLSP , $\widetilde \chi_1^0 \to \bar\nu_i\widetilde \nu_R^1$, 
and  $\widetilde \chi_1^0 \to \bar\nu_i\widetilde \nu_R^2$, respectively. 
As can be seen from Table \ref{relics}, the dominant contributions are coming from 
$\Omega_{\widetilde \nu_R^2}^{\tt ce}$ $(39.7\%)$ and $\Omega_{\widetilde \nu_R^2}^{\tt fo}$ $(59.4\%)$ 
and only $\sim 1\%$ comes from decays to the LSP, as expected. From 
(\ref{sneutrino_angle}),  the mixing angle is inversely proportional to 
$m_{\widetilde \nu_{L L}^i}^{2}-m_{\widetilde \nu_{R R}^i}^{2}$. Thus, as the mass difference between 
left and right handed sneutrinos becomes smaller, their mixing becomes larger. From the fact that 
$m_{\widetilde \nu_R^1}=100 {\rm GeV}$ and $m_{\widetilde \nu_R^2}=1200 {\rm GeV}$ as well as 
$m_{\widetilde \nu_L^i}=2000 {\rm GeV}$, the 
NLSP $\widetilde \nu_R^2$  mixes largely with left-handed sneutrino fields and becomes most likely to 
be produced through the decays in (\ref{channels}). The total  relic density is 0.1036, which lies in 
the $2\sigma$ of the WMAP value (with those from the Sloan Digital Sky Survey) \cite{Spergel:2006hy}:
\begin{eqnarray}
 \Omega_{DM} h^2 = 0.111^{+0.011}_{-0.015}
\end{eqnarray}
Having shown that the parameter set of (\ref{inputs}) can reproduce the required relic density,  
we next investigate the positron flux and discuss the PAMELA, ATIC and FERMI-LAT data 
for the same parameter set.

%%%%%%%%%%%%%%%%%%%%%%%%%%%%%%%%%%%%%%%%%%%%%%%%%%%%%%%%%%%%%%%%%%%%
\subsection{Understanding the PAMELA, ATIC and FERMI-LAT data}
%%%%%%%%%%%%%%%%%%%%%%%%%%%%%%%%%%%%%%%%%%%%%%%%%%%%%%%%%%%%%%%%%%%%

Unlike the decays of heavy superpartners, 
the decay of the NLSP
into the LSP proceeds much more slowly as seen from Table \ref{widths} (see also \cite{maxim}).
The reason is as follows.  As gauge singlets, the sneutrinos  do not couple 
to gauginos but  only to Higgsinos, via
\begin{eqnarray}
\frac{1}{M_R} \left[S \left(\nu_{L} \widetilde{H}_u^0 - e_L \widetilde{H}_u^+\right) + 
H_u^0 \nu_L \widetilde{S} - H_u^+ e_L \widetilde{S}\right] {\bf Y_{\nu}} \tilde \nu_{R},\nonumber\\
\end{eqnarray}
in which the Yukawa interactions of the Higgsinos are
\begin{eqnarray}
&& h_{\mu} S \left(\widetilde{H}_u^0 \widetilde{H}_d^0 - \widetilde{H}_u^+ \widetilde{H}_d^-\right) +
h_{\mu}\left(H_u^0 \widetilde{S} \widetilde{H}_d^0 - H_u^+ \widetilde{S}\widetilde{H}_d^-\right)\nonumber\\
&+& h_{\mu} \left(\widetilde{S} \widetilde{H}_u^0 H_d^0 - \widetilde{S} \widetilde{H}_u^+ H_d^-\right).
\end{eqnarray}
These interaction terms show that the exchanges
of $\widetilde{H}_u^0$ and $\widetilde{S}$ induce
\begin{eqnarray}
\widetilde{\nu}^{2}_R \rightarrow \widetilde{\nu}^1_R \nu_{i} \overline{\nu}_{j},
\end{eqnarray}
 and the $\widetilde{H}_u^+$ exchange gives rise to the decay
\begin{eqnarray}
\label{leptonic}
\widetilde{\nu}^{2}_R \rightarrow \widetilde{\nu}^1_R \ell^\pm_{i} \ell^\mp_{j}.
\end{eqnarray}
The rate of this dileptonic decay is given by
\begin{eqnarray}
\label{rate}
\Gamma_{\ell^\pm_i \ell^\mp_j}
&=&\frac{1}{(2\pi)^{3}}   \left({\overline{\bf Y}}_{\nu}^{\dagger} {\overline{\bf Y}}_{\nu}\right)_{1 1}
\left({\overline{\bf Y}}_{\nu} {\overline{\bf Y}}_{\nu}^{\dagger}\right)_{2 2} \left(\frac{C_a}{\sin\beta}\right)^4  \nonumber\\ &\times&
\frac{m_{\widetilde{\nu}^{2}_{R}}}{32} \left(\frac{m_{\widetilde{\nu}^{2}_{R}}}{m_{\widetilde{\chi}^{+}_{a}}}
\right)^{4} {\mathcal{G}}_{\ell} \left(\frac{m^2_{\widetilde{\chi}^+_{a}}}{m^2_{\widetilde{\nu}^{2}_{R}}},
\frac{m^2_{\widetilde{\nu}^{1}_{R}}}{m^2_{\widetilde{\nu}^{2}_{R}}}\right),
\end{eqnarray}
after summing over all three lepton generations.  Integrating
over the Dalitz density gives
\begin{eqnarray}
{\mathcal{G}}_{\ell}(x,y) &=& x \left(y-1\right) \left[ x
\left(5-6 x\right) + y \left(5 x
-2\right)\right] \nonumber\\ &-& 2 y^2 \log y - \Big[2
\left(y - x\right) \left(x -1\right)\nonumber\\
&\times& \left(y + x + x y  - 3 x^2 \right) \log
\left(\frac{x-y}{x-1}\right)\Big] ,
%\nonumber\\
%%&=&  \frac{1}{6}\left(1-8 y - 12 y^2 \log y + 8 y^3 - y^4\right)\nonumber\\ &-& \frac{2}{15 x} (y -1)^3 + {\cal{O}}\left(\frac{1}{x^2}\right)
\end{eqnarray}
which appears in ~(\ref{rate}) with the indicated arguments.
The two chargino states $\widetilde{\chi}^+_{a}$ ($a=1,2$) with
masses $m_{\widetilde{\chi}^+_a}$, also appear in the decay
rate via $\widetilde{H}^+_{u} = C_1 P_R \widetilde{\chi}_1^+ +
C_2 P_R \widetilde{\chi}_2^+$, in which
\begin{eqnarray}
\frac{C_1 C_2}{C_2^2 - C_1^2} = \frac{r_{W 2} (\sin\beta + r_{\mu 2} \cos\beta)}{1-r_{\mu 2}^2 + r_{W 2}^2 \cos 2\beta}
\end{eqnarray}
where $C_2 = \sqrt{1-C_1^2}$, $r_{W 2} = \sqrt{2} M_W/M_2$,
$r_{\mu 2} = h_{\mu} \langle S \rangle / M_2$, and $M_2$ is the
$SU(2)_{L}$ gaugino mass.

Various analyses \cite{decay} of PAMELA and other satellite data
suggest that, rather model
independently, the DM candidate must have a lifetime of $\sim 10^{26}\ {\rm sec}$.
%This time scale is obviously longer than the age of the universe by several orders of magnitude.
%As previously mentioned, the cosmic ray data can be interpreted to imply an absolutely 
%stable DM particle (e.g. $\widetilde{\nu}^1_R$) with a 
%long-lived particle (e.g. $\widetilde{\nu}^{2}_{R}$) with a mass similar to 
%the DM candidate, such that its decays into the DM candidate reproduces the observations.
This is consistent with the dileptonic decay rate ~(\ref{rate}) in that for the parameter 
choice in (\ref{inputs}), one finds
\begin{eqnarray}
\label{ratex}
\Gamma_{\ell^\pm_i  \ell^\mp_j} = 6.8\times 10^{-51}\ {\rm GeV},
\end{eqnarray}
leading to a $\sim 10^{26}\ {\rm sec}$ lifetime. This arises from the small 
$\langle S \rangle/M_{GUT}$ ratio
that sets the neutrino mass scale. One notices that the value of the rate involves 
the sum of all three generations of charged leptons ($i,j=1,2,3$) as well as all the 
conjugated modes. Note that there is no contribution to the positron production 
from the third right-handed sneutrino since it only lives $10^{-5}$ sec and thus decays much faster. 

In contrast to the case of the MSSM with right-handed neutrinos
\cite{moroi,maxim}, the $\mu$ term and the neutrino Yukawa couplings are induced 
dynamically in our model. The two models also differ in terms of the predicted 
correlation between the neutrino masses~(\ref{mneut}) and the dileptonic 
decay rate~(\ref{rate}). In the MSSM, these quantities are directly
correlated via the Yukawa couplings ${\overline{\bf Y}}_{\nu} \sim {\cal{O}}(10^{-13})$. 
In our model, the rate involves both $\langle S
\rangle /M_{GUT}$ and $m_{\widetilde{\nu}^2_R}/m_{\widetilde{H}_u}$, and, in the heavy 
Higgsino regime where $m_{\widetilde{\chi}_2^{+}} \simeq m_{\widetilde{H}_u} \simeq \mu \gg M_2, M_W$, 
it scales as $(m_{\widetilde{\nu}^2_R}/m_{\widetilde{\nu}^1_R})^4$, not as $(\langle S \rangle /M_{GUT})^4$. 
Hence, in the large $\mu$ limit there is only a mild dependence on $m_{\widetilde{H}_u}$,
and the ``see-saw scale" that suppresses the neutrino masses is then independent of the 
scale that governs the NLSP decays. However, in this regime the DM lifetime is longer 
than $10^{26}\ {\rm sec}$, and hence the heavy Higgsino regime is not particularly 
preferred given present data.

We now crystallize the qualitative approach presented above with a more detailed calculation 
of the positron flux originating from the decay (\ref{leptonic}) with the decay width (\ref{rate}). 
We  follow the procedure in \cite{decay}, to which we refer the reader for further details. 
The source term of the diffusion equation is
\begin{eqnarray}
 Q_{\ell^\pm_i}(E_{\ell^\pm_i},{\bf r})=\frac{n_{\rm NLSP}}{\tau_{\rm NLSP}}\,\frac{dN_{\ell_i^\pm}}{dE_{\ell^\pm_i}},
\end{eqnarray}
where $n_{\rm NLSP}$ is the number density of the NLSP, $\tau_{\rm NLSP}$ is the 
lifetime (which will be set as a free parameter to be fit), and $dN_{\ell^\pm_i}/dE_{\ell^\pm_i}$ is 
the energy distribution of $\ell^\pm_i$ from the decay 
$\widetilde{\nu}^{2}_R \rightarrow \widetilde{\nu}^1_R \ell^\pm_{i} \ell^\mp_{j}$. Before proceeding to evaluate 
 $dN_{\ell^\pm_i}/dE_{\ell^\pm_i}$,  we first  determine $\tau_{\rm NLSP}$.

Since both the  LSP and the NLSP are  components of the DM, other particles decay to either of them. 
Then, the ratio of the number densities of the LSP and NLSP should be equal to the ratios of the 
branching rates into either state. Explicitly, 
\begin{eqnarray}
\displaystyle \frac{n_{\rm LSP}}{n_{\rm NLSP}} = \frac{\sum_{A,B}{\cal B}(A\to B\, \widetilde\nu_R^1)}
{\sum_{A^\prime,B^\prime}{\cal B}(A^\prime\to B^\prime\, \widetilde\nu_R^2)}\equiv \alpha
\label{ndensity_1}
\end{eqnarray}
where $A,B,A^\prime$, and $B^\prime$ represent all possible particles of the model and the right hand 
side is known once we fix the free parameters. For example, numerical evaluation yields, 
$\sum_{A,B}{\cal B}(A\to B\, \widetilde\nu_R^1)=0.0718$ and 
$\sum_{A,B}{\cal B}(A\to B\, \widetilde\nu_R^2)=0.428$ for the particular parameter set considered here. 
There are two unknown in the left hand side of (\ref{ndensity_1}). The second relation needed arises from the mass density of the DM,  $\rho_{\rm DM}(r)$, as
\begin{eqnarray}
m_{\widetilde \nu_R^1}\, n_{\rm LSP} + m_{\widetilde \nu_R^2}\, n_{\rm NLSP} = \rho_{\rm DM}(r)
\label{ndensity_2}
\end{eqnarray}
where $\rho_{\rm DM}({\bf r})$ is further parametrized by adopting the Navarro-Frank-White 
density profile \cite{Navarro:1996gj}. For the diffusion parameters, we use the three sets 
from models M1, M2 and MED \cite{decay}. 

We return to the evaluation of  the $dN_{\ell^\pm_i}/dE_{\ell^\pm_i}$ term. The definition  and normalization of this term depends on whether the 
positron/electron pair is produced directly from the decay 
$\widetilde{\nu}^{2}_R \rightarrow \widetilde{\nu}^1_R e^- e^+$ (that is, $i=j=1$)
or through some cascade decays. Our numerical analysis shows that in addition to the direct $e^- e^+$
 production case, the $e^\pm\mu^\mp$ has also significant decay width. In this case the muon 
$\mu^\pm$ will decay further as  $\mu^\pm\to \nu_\mu e^\pm \nu_e$ where the appropriate combinations of 
$\nu_\mu$ and $\nu_e$ should be understood based on the charge of $\mu$ or $e$. In principle, the $\tau$ lepton
can also be produced directly which then requires two cascade decays to get the positron/electron pair. So, we  consider direct productions of positron/electron, or indirect production only through the muon cascade. The details of how to treat the 
cascade productions of positron/electron from the muon cascade can be found for example in \cite{Chen:2009mj}. If the 
electron/positron pair is produced directly, then the $dN_{\ell^\pm_i}/dE_{\ell^\pm_i}$ is given
\begin{eqnarray}
 \frac{dN^{[e^-e^+]}_{e^\pm}}{dE_{e^\pm}}=\frac{1}{\Gamma_{e^+e^-}}\int dE_{e^\mp} \frac{d^2\Gamma}{dE_{e^+}dE_{e^-}}.
\label{direct}
\end{eqnarray}
where $\Gamma_{e^+e^-}$ is the decay width for the direct electron/positron production case. 
If positron/electron is produced from the decay 
$\widetilde{\nu}^{2}_R \rightarrow \widetilde{\nu}^1_R e^\pm \mu^\mp$ followed by the decay of the muon 
$\mu^\mp\to \nu_\mu e^\mp \nu_e$, we write the $dN/dE$ term 
for the positron only for notational simplicity (the electron case is very similar): 
\begin{eqnarray}
 \frac{dN^{[e^\pm \mu^\mp]}_{e^+}}{dE_{e^+}}=\frac{1}{N_{e^\pm\mu^\mp}}
\left( \frac{d\Gamma^{e^+\mu^-}}{dE_{e^+}} 
+ \int dE_{\mu^+} \frac{d^2\Gamma^{e^-\mu^+}}{dE_{e^+}dE_{\mu^+}} \right)
\label{cascade}
\end{eqnarray}
where the second term is represented generically after carrying out the integral over 
the electron energy  $E_{e^-}$. In this term, the muon decays further to a positron with energy $E_{e^+}$. The combined contribution is included in the second term. The first term corresponds to the case where the leptons in the final state are $e^+\mu^-$ so that there is no further decay considered 
for the positron case.  The normalization factor $N_{e^\pm\mu^\mp}$ is given by  
\begin{eqnarray}
N_{e^\pm\mu^\mp} = \Gamma_{e^+\mu^-} 
+ \int dE_{e^+} dE_{\mu^+} \frac{d^2\Gamma^{e^-\mu^+}}{dE_{e^+}dE_{\mu^+}}
\end{eqnarray}

The solution of the diffusion equation for the flux $\Phi_{e^\pm}$ is given in terms of the 
Green's function in \cite{decay} as
\begin{eqnarray}
 \Phi_{e^\pm}(E_{e^\pm}) = \frac{c}{4\pi\, m_{\rm eff} 
\tau_{\rm NLSP}}\int dE\, G(E,E_{e^\pm})\,\frac{dN_{e^\pm}}{dE}
\label{ratio}
\end{eqnarray}
where either (\ref{direct}) or (\ref{cascade}) should be used for the $dN_{e^\pm}/dE$ term depending on the relevant 
decay mode. Here $m_{\rm eff}=\alpha\, m_{\widetilde{\nu}^1_R} + m_{\widetilde{\nu}^2_R}$ is an 
effective mass term originating from the solution of (\ref{ndensity_1}) and (\ref{ndensity_2}). 
We further approximate the Green's function as in \cite{Ibarra:2008qg}  to maximize computational efficiency. 

The flux $\Phi_{e^\pm}$ obtained  as a solution of the diffusion equation is then used to 
define the ratio for the positron flux $R_{e^\pm}$:
\begin{eqnarray}
 R_{e^\pm} = \frac{\Phi^{\rm tot}_{e^+}}{\Phi^{\rm tot}_{e^-} + \Phi^{\rm tot}_{e^+}},
 \label{ratio-positron}
\end{eqnarray}
where $\Phi^{\rm tot}_{e^\pm}$ is the sum of $\Phi_{e^\pm}$ and the background flux in our 
galaxy and we define $\Phi^{\rm tot}_{e^\pm}$
\begin{eqnarray}
&& \Phi^{\rm tot}_{e^-}=\Phi_{e^-}+ \kappa^-\, (\Phi^{\rm prim}_{e^-} 
+ \Phi^{\rm sec}_{e^-})\,,\nonumber\\
&& \Phi^{\rm tot}_{e^+}=\Phi_{e^+} +  \kappa^+\, \Phi^{\rm sec}_{e^+}\,,
\end{eqnarray}
where $\kappa^{\pm}$ represents the uncertanities in the 
$e^+(e^-)$ background. $(\kappa^-,\kappa^+)=(0.7,0.9)$ is used
 in the numerical study. The following electron and 
positron backgrounds $(\Phi^{\rm prim}_{e^-},
\Phi^{\rm sec}_{e^\mp})$ 
in our galaxy are approximated as \cite{Baltz:1998xv} 
in units of $\rm (GeV\,cm^2\,sec\,sr)^{-1}$
\begin{eqnarray}
&&\Phi^{\tt prim}_{e^-}=\frac{0.16E^{-1.1}}{1+11E^{0.9}+3.2E^{2.15}}\nonumber\\
&&\Phi^{\tt sec}_{e^-}=\frac{0.7E^{0.7}}{1+110E^{1.5}+600E^{2.9}+580E^{4.2}}\nonumber\\
&&\Phi^{\tt sec}_{e^+}=\frac{4.5E^{0.7}}{1+650E^{2.3}+1500E^{4.2}}
\end{eqnarray}

The quantity $R_{e^\pm}$ in (\ref{ratio-positron}) can be used to compare with the PAMELA data. Similarly the total 
flux $\Phi^{\rm tot}_{e^-} + \Phi^{\rm tot}_{e^+}$ scaled by $E_{e^\pm}^3$ is the relevant observable for 
the ATIC and the FERMI-LAT experiments.
\begin{figure}[htb]
%\vskip 0.2in
\begin{center}
\hspace*{-0.4cm}
	\includegraphics[width=3.6in,height=2.4in]{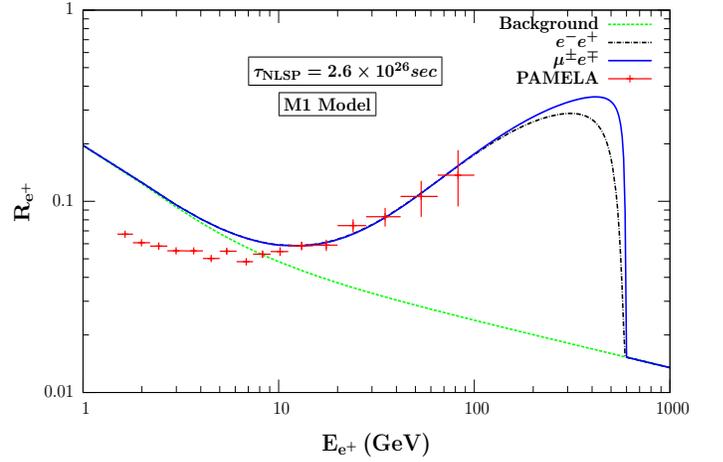}
\end{center}
\vskip -0.2in
\caption{The positron flux $R_{e^+}$ as a function of positron energy in the M1 
propagation model.  The direct positron channel ($e^-e^+$) and the indirect one through 
muon decay ($\mu^\pm e^\mp$) are shown separately. 
The background and the PAMELA data are also shown. 
The fitted lifetime of NLSP is also indicated.}\label{flux1}
\end{figure}
\begin{figure}[htb]
%\vskip 0.2in
\begin{center}
\hspace*{-0.4cm}
	\includegraphics[width=3.6in,height=2.4in]{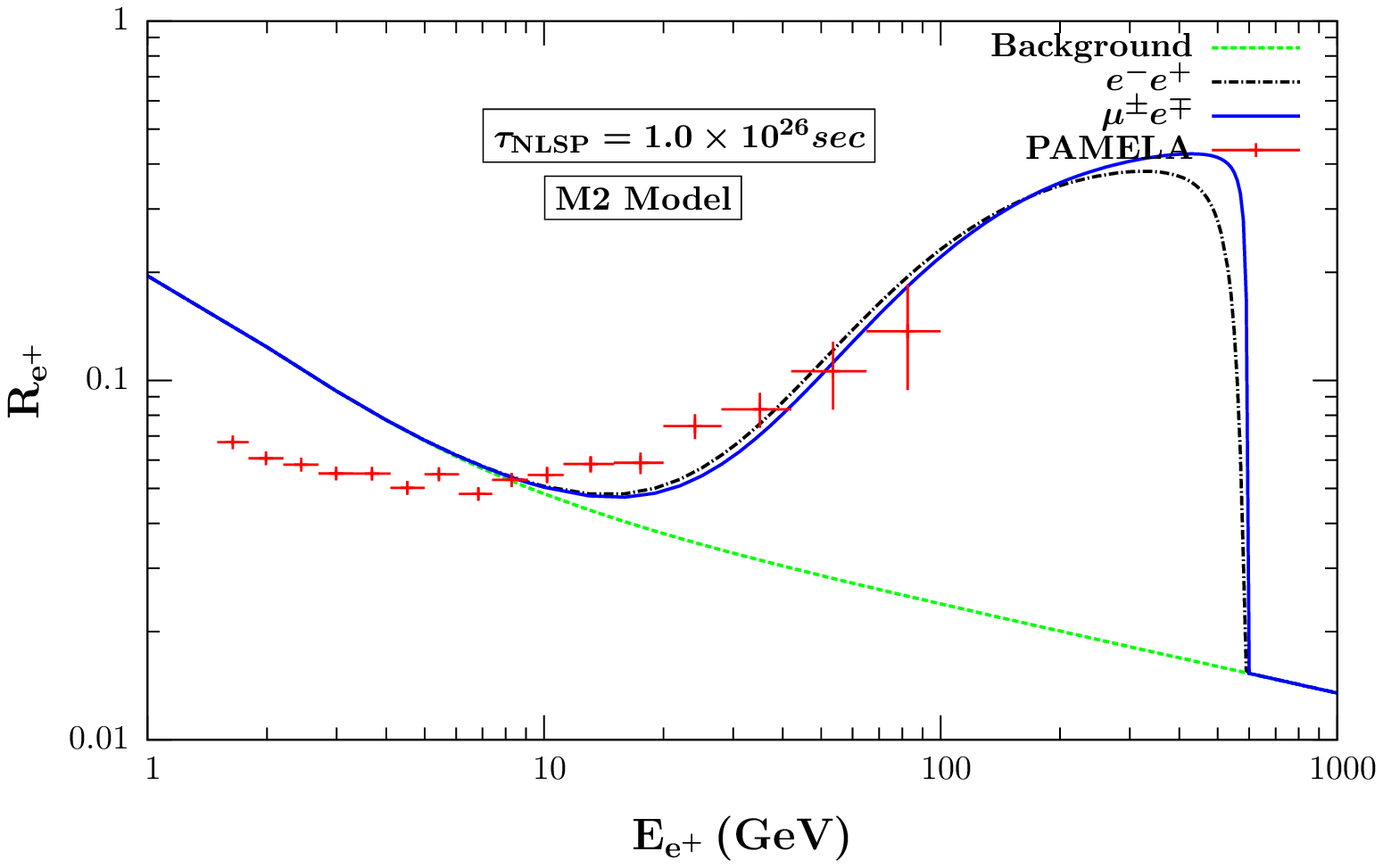}
\end{center}
\vskip -0.2in
\caption{The positron flux $R_{e^+}$ as a function of positron energy in the M2 
propagation model. The direct positron channel ($e^-e^+$) and the indirect one through 
muon decay ($\mu^\pm e^\mp$) are shown separately.
The background and the PAMELA data are also shown. 
The fitted lifetime of NLSP is also indicated.}\label{flux2}
\end{figure}
\begin{figure}[htb]
%\vskip 0.2in
\begin{center}
\hspace*{-0.4cm}
	\includegraphics[width=3.6in,height=2.4in]{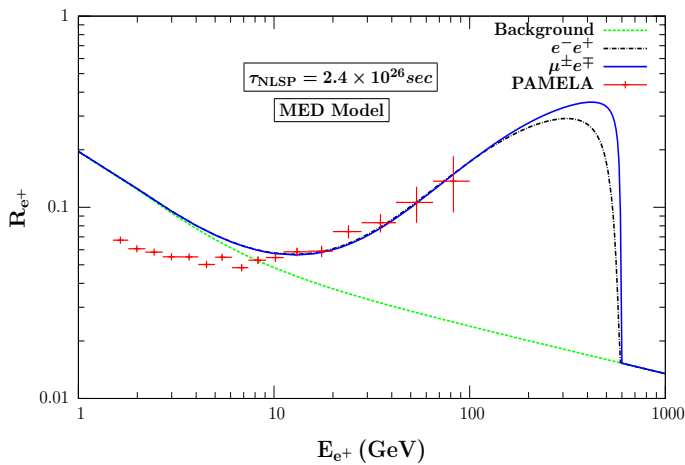}
\end{center}
\vskip -0.2in
\caption{The positron flux $R_{e^+}$ as a function of positron energy in the MED 
propagation model. The direct positron channel ($e^-e^+$) and the indirect one through 
muon decay ($\mu^\pm e^\mp$) are shown separately. 
The background and the PAMELA data are also shown. 
The fitted lifetime of NLSP is also indicated.}\label{flux3}
\end{figure}

Our results for the positron flux are shown in Figs.~(\ref{flux1}), (\ref{flux2}), 
and (\ref{flux3}) 
as the ratio $R_{e^+}$ as a function of the positron energy 
$E_{e^+}$ for each propagation model M1, M2, and MED, respectively.  
The direct and indirect production 
of the positron as well as the expected background are shown separately. 
The figures show that the models M1 and MED 
fit much better than the M2 model. The fitted curves significantly 
deviate from the PAMELA data 
(shown in the figures) for energies less than about $5\ {\rm GeV}$ but 
are consistent with the experimental data for $E>5\ {\rm GeV}$.  
The fitted values of the lifetimes 
are  ${\cal O}(10^{26}\ {\rm sec})$,  which are in the expected range. 
The lifetimes for the M1 and MED models 
are close to each other, and the value is only slightly smaller for the M2 model. Note also that the 
constraint on the lifetime of the decaying DM from the cosmic microwave background 
(CMB) is analysed in 
\cite{Zhang:2007zzh}, and our results obey the lower bounds. 
The positron emitted through 
the $\mu^\pm e^\mp$ channel deviates only from the direct channel part 
of the high energy tail.

\begin{figure}[t]
%\vskip 0.2in
\begin{center}
\hspace*{-2.0cm}
	\includegraphics[width=4.2in,height=2.4in]{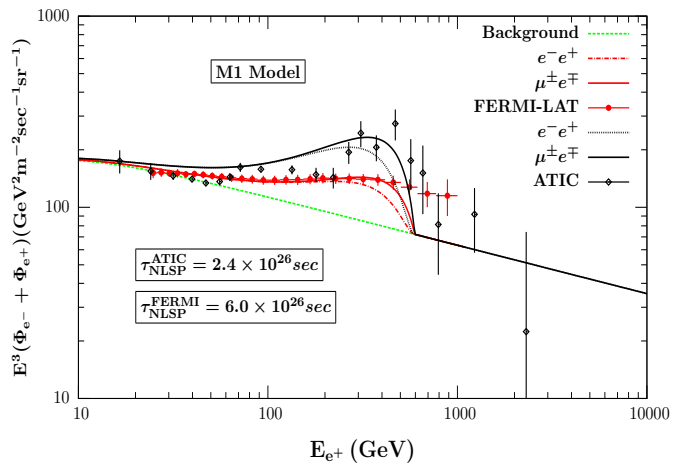}
\end{center}
\vskip -0.2in
\caption{The total flux scaled by $E_{e^\pm}^3$ as a function of the positron
 energy in 
the M1 propagation model. Both the direct positron channel ($e^-e^+$) and the 
indirect one through 
muon decay ($\mu^\pm e^\mp$) are shown separately. The background and the 
ATIC and FERMI-LAT 
 data are also shown. 
The fitted lifetimes of NLSP for each data are also indicated.}\label{atfer1}
\end{figure}

\begin{figure}[b]
%\vskip 0.2in
\begin{center}
\hspace*{-2.0cm}
	\includegraphics[width=4.2in,height=2.4in]{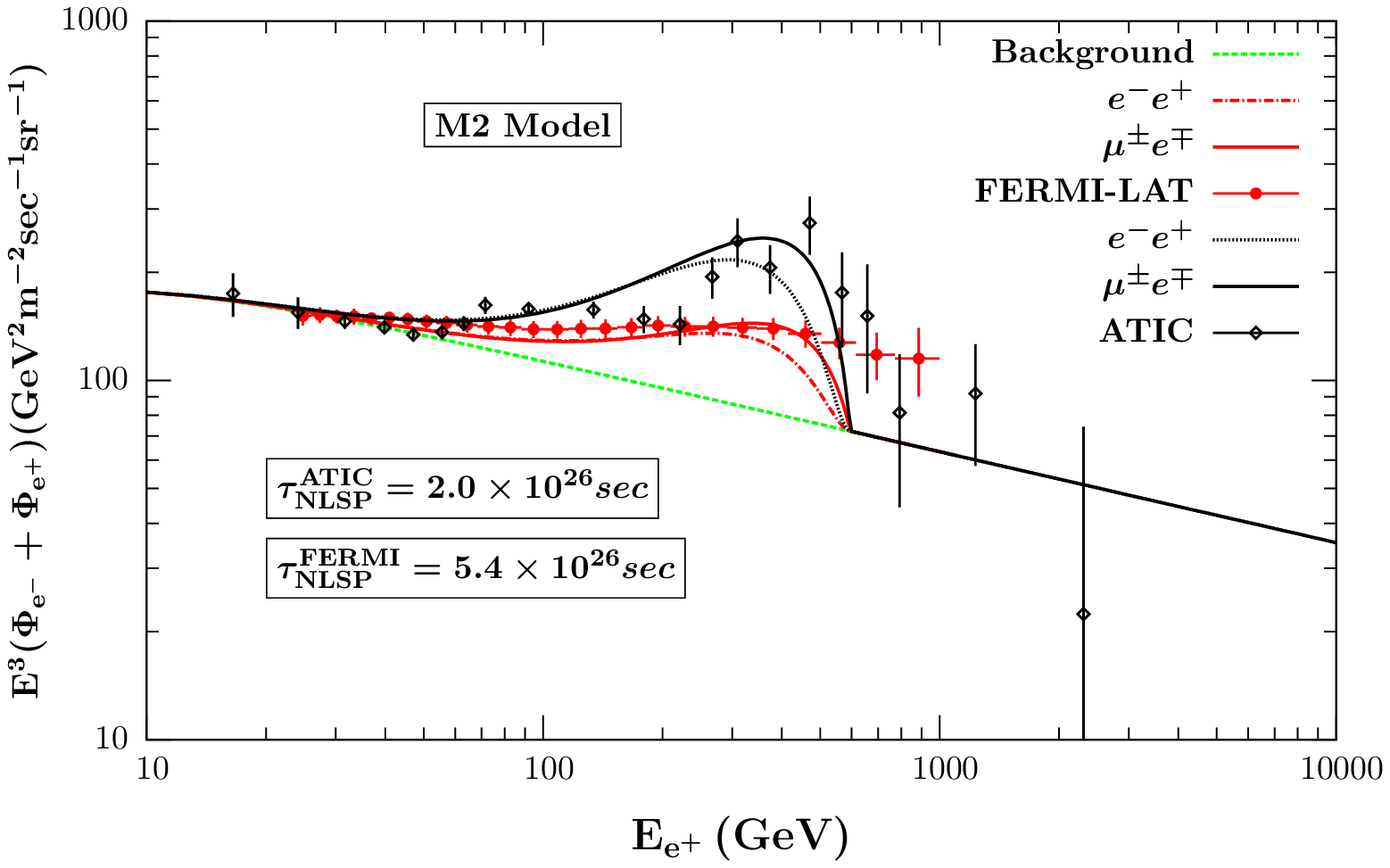}
\end{center}
\vskip -0.2in
\caption{The total flux scaled by $E_{e^\pm}^3$ as a function of the 
positron energy in  
the M2 propagation model. Both the direct positron channel ($e^-e^+$) 
and the indirect one through 
muon decay ($\mu^\pm e^\mp$) are shown separately. The background and 
the ATIC and FERMI-LAT 
 data are also shown. 
The fitted lifetimes of NLSP for each data are also indicated.}\label{atfer2}
\end{figure}

\begin{figure}[htb]
%\vskip 0.2in
\begin{center}
\hspace*{-2.0cm}
	\includegraphics[width=4.2in,height=2.4in]{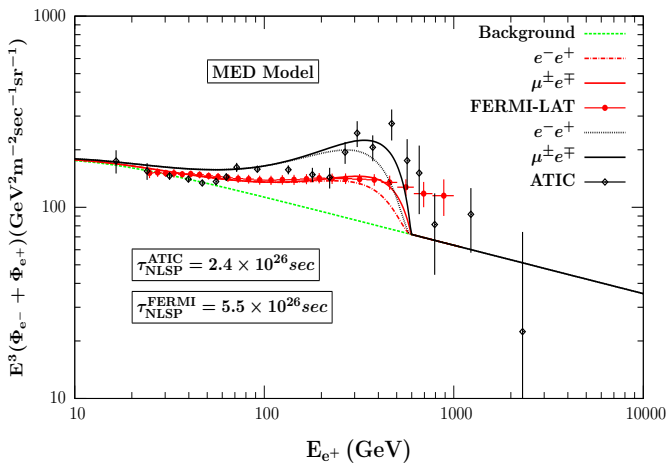}
\end{center}
\vskip -0.2in
\caption{The total flux scaled by $E_{e^\pm}^3$ as a function of the 
positron energy in  
the M2 propagation model. Both the direct positron channel ($e^-e^+$) 
and the indirect one through 
muon decay ($\mu^\pm e^\mp$) are shown separately. The background and 
the ATIC and FERMI-LAT 
 data are also shown. 
The fitted lifetimes of NLSP for each data are also indicated.}\label{atfer3}
\end{figure}

Similar graphs are shown in Figs.~(\ref{atfer1}), (\ref{atfer2}), and (\ref{atfer3}) 
for the total flux scaled by the positron energy $E_{e^\pm}^3$ for 
the propagation models M1, M2, and MED, 
respectively. Again the direct and indirect production 
of the positron as well as the expected background are shown separately. 
The ATIC and the FERMI-LAT data 
are also included. As seen from the figures,  our signal is a 
reasonably good fit to the ATIC and 
FERMI data. Our signal explains the FERMI-LAT data very well 
up to $\sim 400$ GeV for the $e^\pm e^\mp$ case and gets even 
better for 
the $\mu^\pm e^\mp$ case, where the signal is consistent up to $\sim 500$ GeV. 
In the ATIC case, the signal explains the low energy as well as the peak regions 
better than the intermediate enery interval. 
Once again, considering the $\mu^\pm e^\mp$ mode makes the situation better in 
the end-region.
In either case (ATIC or FERMI-LAT), the signal would fit better especially in the 
high energy region for another set of parameter values. For example, 
a bigger gap between the LSP and NLSP mass would allow more energetic 
$e^\pm$ which would shift the dying tail of the signal to the right.

Unlike the PAMELA case, where M1 and MED scenarios 
are favored, the M2 model works better for ATIC, while the FERMI-LAT  data 
slightly favors the M1 
and MED models.  The fitted lifetimes 
for the ATIC data are very close  
for the propagation models and are 
also quite consistent with the ones 
for the PAMELA case (especially for the M1 and MED models). On the other hand, 
the lifetimes for 
the FERMI-LAT data are the same for the three models considered (the fitted 
curves are less sensitive 
to the types of the model used as long as the lifetimes are in 
the $(5.4-6.0)\times10^{26}$ sec range). For the M1 and MED models, 
the FERMI-LAT data requires lifetimes about twice bigger than than the ones for 
the ATIC and PAMELA cases. The ratio is bigger for the M2 propagation model.


From the results, we can conclude that it is possible to 
satisfactorily explain the PAMELA+ATIC data simultaneously 
(except the data in the 100-200 GeV range for the ATIC case) with a 
 consistent lifetime for the DM.  It is also possible to explain the FERMI-LAT data, which requires a slightly larger lifetime for the DM and works well for energies up to 500 GeV. 

Having discussed the electron/positron excess as well as the total flux, 
we now consider briefly 
the issue that no significant
excess was observed in the anti-proton flux \cite{pamela2}. Unlike the
 case of the leptonic decays, 
the hadronic decays are generated only through the $F$-term interactions 
in the superpotential. 
The relevant interaction terms in the Lagrangian are given by
\begin{eqnarray}
\label{lagran}
{\cal{L}} \sim \left({\overline{\bf Y}}^{\dagger}_{\nu} 
{\overline{\bf Y}}_{\nu}\right)_{2 1} 
\left( \frac{v}{\sin\beta} \right) \left({\cal R}^{-1}\right)_{u i} H_i^0
 \widetilde{\nu}^{2 \star}_{R} 
\widetilde{\nu}^{1}_{R} + \mbox{H. c.},
\end{eqnarray}
in which $H_i^0$ ($i=1,\dots,3$) are the neutral CP-even Higgs
bosons and  ${\cal{R}}_{u i}$ is the fraction of $H_i^0$ in
$H_{u}^0$ \cite{lisa}. Higgs bosons $H_i^0$ produced in the
$\widetilde{\nu}^2_{R}$ decay \cite{maxim} further decay into
pairs of quarks (including top quark pairs as suggested by the
mass spectrum in (\ref{inputs}) and in Table \ref{spectrum}, gauge bosons, 
and Higgs
bosons. The decay rate to quarks is
\begin{eqnarray}
\label{rateq}
\Gamma_{q {\bar q}} &=& \frac{N_C}{(2 \pi)^3} \left({\overline{\bf Y}}^{\dagger}_{\nu} {\overline{{\bf Y}}}_{\nu}\right)_{1 2}
\left({\overline{\bf Y}}^{\dagger}_{\nu} {\overline{{\bf Y}}}_{\nu}\right)_{2 1}
\sum_{i=1}^3\left( \frac{{\cal{R}}_{u i}}{\sin \beta}\right )^4\nonumber\\
& \times & \frac{m_{{\widetilde \nu}_R^2}}{32} \left(\frac{m_q}{m_{{\widetilde \nu}_R^2}}\right)^2\ 
{\mathcal{G}}_{q} \left(\frac{M^2_{H_i^0}}{m^2_{\widetilde{\nu}^{2}_{R}}},
\frac{m^2_{\widetilde{\nu}^{1}_{R}}}{m^2_{\widetilde{\nu}^{2}_{R}}}\right).
\end{eqnarray}
The integrated Dalitz density has the functional form
\begin{eqnarray}
{\mathcal{G}}_q (x,y) &=& -4 \left(1-y\right) + \alpha \log{y} + \Bigg[\frac{1}{\beta}\left(2 \beta^2 - \gamma \right)\nonumber\\ &\times&
\log\left(\frac{\gamma + (1-y) \beta }{\gamma -(1-y) \beta}\right)\Bigg]
\end{eqnarray}
with $\alpha = - 2 x + (1+y)$, $\beta^2 = -x^2 - x \alpha +
\gamma$, and $\gamma= (1-y)^2 - x (1+y)$.

The light quarks are produced directly or indirectly through heavy quark, 
gauge boson, 
and Higgs boson decays, and hadronize to produce protons and anti-protons. 
An inspection of 
~(\ref{rateq}) and ~(\ref{rate}) shows that the anti-proton signal is 
suppressed compared 
to the positron signal by a kinematic factor of order 
$\sim 10 m^2_q/m^2_{\widetilde{\nu}_R^2}$, 
which is ${\cal O}\left(10^{-10}\right)$ for $u,d$ quarks and 
${\cal O}\left(10^{-4}\right)$ for $b$ quarks, 
as needed for consistency with PAMELA~\cite{pamela2}.

However, the Higgs bosons that are produced in NLSP to LSP decays 
fragment efficiently into gauge 
bosons and top quarks, which in turn give off light quarks and 
anti-quarks with no apparent 
Yukawa suppression. Moreover, the leptons produced with the
rate given in ~(\ref{rate}) can give rise to hadronic final 
states at the loop level. 
These indirect contributions are expected to yield anti-protons
 with an efficiency around $10 \%$ \cite{maxim,antiprot}.

The needed suppression of the anti-proton flux with respect to 
the positron flux can stem
from various effects, such as the interaction strengths 
in ~(\ref{lagran}).  One way to 
suppress $\Gamma_{q\bar{q}}$ with respect to $\Gamma_{l^+l^-}$ 
is to have  \cite{maxim}
\begin{eqnarray}
\label{pbound}
{\left|\left({\bf Y^{\dagger}_{\nu}} {\bf Y_{\nu}}\right)_{2 1} \right|^2} \simlt
10^{-4}\ {\left({\bf Y_{\nu}^{\dagger}} {\bf Y_{\nu}}\right)_{1 1}
\left({\bf Y_{\nu}} {\bf Y_{\nu}^{\dagger}}\right)_{2 2}}
\end{eqnarray}
which imposes an overall suppression on the flavor-changing entries of the 
neutrino Yukawa matrices with respect to the flavor-conserving
 entries. A justification or 
realization of such a structure would in principle require a 
detailed knowledge of the flavor 
structure of ~(\ref{mneut}).  However, it is worth noting that
 there is no direct correspondence 
in general between the right-handed neutrino and the active 
(left-handed) neutrino sectors, 
which can lend credence for the needed suppression.

%%%%%%%%%%%%%%%%%%%%%%%%%%%%%%%%%%
\section{Conclusions}
%%%%%%%%%%%%%%%%%%%%%%%%%%%%%%%%%%

We have presented a simple extension of
the MSSM with an additional $U(1)^\prime$ gauge symmetry that
couples predominantly to Higgs fields.  While other sneutrino 
dark matter models have 
been proposed to explain the cosmic ray observations, the model 
presented here does so
while also resolving the naturalness problems of the MSSM. In 
this model, an electroweak 
scale $\mu$ term and appropriately suppressed Dirac neutrino 
masses are generated upon 
$U(1)^\prime$ breaking. The right-handed
sneutrinos $\widetilde{\nu}^1_R$ and  $\widetilde{\nu}^2_R$ 
are the lightest and next-to-lightest
superpartners, allowing for a natural explanation of PAMELA and
other experiments like ATIC and FERMI-LAT in the context of sneutrino dark matter. 

A complete study of the DM relic density in this model includes contributions 
 from the LSP and the NLSP, which is essentially stable with respect to the lifetime of the universe. We also 
analysed the contributions coming from late decaying particles, 
{\it i.e.}, particles that decay after 
freeze-out. For the parameter set in (\ref{inputs}), we obtain 
$0.1036$ for the total relic density of 
right-handed sneutrino,  consistent with the current WMAP value.

We then discussed the PAMELA, ATIC and FERMI-LAT data,  considering 
$\widetilde \nu_R^2\to \widetilde \nu_R^1 l_i^\pm \l_j^\mp$ as the 
source decay for the observed positron excess. The possibility 
of producing a final positron through muon is also discussed separately. The suppression of the anti-proton flux compared to the positron flux suggests the model-building constraint that the flavor-changing entries of the neutrino Yukawa coupling matrix are suppressed with respect to the flavor-conserving entries.
The fitted lifetimes for the DM, of the order of $10^{26}$ sec, fit the PAMELA data well for positron energies greater than  $5$ GeV, as well as the ATIC and FERMI-LAT cases. The fit 
is better for the M1 and MED models described in the text. It can be concluded 
from the values of the fitted lifetimes that it maybe possible to explain 
the PAMELA+ATIC  and FERMI-LAT data simultaneously 
in the corresponding energy ranges, though the FERMI-LAT data requires a 
slightly larger lifetime values for the DM. 
In principle, it is foreseeable that  a better scan of the model 
parameters as well as implementing an improved 
fitting procedure could offer an even closer simultaneous 
explanation of these three sets of experimental data.

%%%%%%%%%%%%%%%%%%%%%%%%%%%%%%%%%%
\section{Acknowledgements} 
%%%%%%%%%%%%%%%%%%%%%%%%%%%%%%%%%%

We thank the referee for helpful comments and feedback, and for pointing out several crucial issues.  IT also thanks Heather Logan for 
useful discussions. The work of DAD is supported by
the Turkish Academy of Sciences via the GEBIP grant and by the
Turkish Atomic Energy Authority via the CERN-CMS Research
Grant.  LLE is supported by the DOE grant
DE-FG-02-95ER40896. The work of MF and IT is supported in part by
the NSERC of Canada under the Grant No. SAP01105354. 
The work of LS is supported in part by a TUBITAK post-doctoral fellowship.

\end{document}